\documentclass[a4paper,12pt]{article}
\usepackage{graphicx,amsmath,amsfonts,amssymb,cite}
\usepackage{bm}
\newcommand{\be}{\begin{equation}}
\newcommand{\ee}{\end{equation}}
\newcommand{\ba}{\begin{eqnarray}}
\newcommand{\ea}{\end{eqnarray}}
\newcommand{\nl}{\nonumber \\}

\begin{document}


\begin{center}
{\Large\bf Cluster duality}
\end{center}

\begin{center}
{S. S. AFONIN}
\end{center}

\begin{center}
{\it V. A. Fock Department of Theoretical Physics, St. Petersburg University,\\
1 ul. Ulyanovskaya, 198504, Russia.\\
}
\end{center}

\begin{abstract}
It is well known that the linear mass spectrum of light mesons in the large-$N_c$ limit
is dual to the perturbative QCD continuum. We find the form of the linear spectrum which is
maximally dual to the perturbation theory. The obtained ansatz turns out to be the spectrum of the
Lovelace-Shapiro dual amplitude. This spectrum is chirally
symmetric in the sense that it corresponds to zero values for the
order parameters of chiral symmetry breaking in QCD.
We further assume that the actual spectrum is dual to the
maximally dual one within the QCD sum rules based on OPE and in this way recover some known
important relations and statements. The experimental spectrum seems to support the assumed duality:
The masses of resonances cluster near certain
equidistant values of energy squared and the positions of clustering follow the maximally dual
ansatz. The latter fact signals the chiral symmetry restoration in
the excited states.
We discuss also two topics of recent interest: nonlinear corrections to the string-like
spectrum and the dimension-two gluon condensate.
\end{abstract}



\section{Introduction}

Thirty years ago A. Migdal~\cite{prog} considered the problem of approximation of the
perturbative asymptotics
for the vector correlator (the so-called parton model logarithm) by an infinite sum of
poles in the
"best possible way". The obtained spectrum happened to be linear for masses, with the corrections
to the perturbative logarithm being exponentially small.
Quite unexpectedly, similar result was recently obtained in the framework of AdS/QCD~\cite{low}.
At present many theoretical and phenomenological indications have been accumulated that with
a good accuracy the spectrum is linear for masses square, {\it i.e.} it has a string-like form.
In addition, according to the Operator Product Expansion (OPE) for two-point
correlators~\cite{svz}, the corrections to the perturbative
logarithm have to decrease polynomially rather than exponentially.
On the other hand, saturating the two-point correlators by
the linear spectrum for masses square one can reproduce
correctly the analytical structure of OPE (polynomial corrections) with non-vanishing residues
(constant in asymptotics). In this respect
an interesting question arises: What linear spectrum for masses square approximates the asymptotics of correlators
in the "best possible way"? Needless to say, the answer to this question is of importance
for the effective string and AdS/QCD approaches. In the given paper we propose a solution of
this problem within the QCD sum rules in the large-$N_c$ limit~\cite{hoof}. We show that in this
case the properties
of linear spectrum and OPE are completely determined by those of Bernoulli polynomials. This very
fact
allows to solve the problem. For the vector mesons the solution turns out to be nothing but the
spectrum of the Lovelace-Shapiro (LS) dual model~\cite{LS}. We demonstrate also that the
deviations from this spectrum
in the axial-vector channel can be ascribed to the spontaneous Chiral Symmetry Breaking (CSB).
We propose a hypothesis that the actual spectrum is dual to the maximally dual one in the sense
that
OPE is identical for both cases. In a particular realization of this concept we rederive some
important relations. The pattern of clustering of the experimental spectrum seems to be tightly
related with the assumed duality.

The paper is organized as follows. In Section~2 the general formalism of QCD sum rules in the
large-$N_c$ limit is presented for the case of linear spectrum. In Section~3 we derive the
maximally dual spectrum. Section~4 is devoted to the analysis of corrections to the linear
spectrum.
Section~5 contains some comments on the sum rules with finite number of states.
Section~6 deals with experimental spectroscopy. Dimension-two gluon condensate is discussed
in Section~7.
Some auxiliary material and discussions are given in Appendices.
We conclude in Section~8.

\section{Sum rules for linear spectrum}

Let us briefly remind the formalism of
QCD sum rules in the large-$N_c$ limit. In this limit the
two-point correlation functions of quark currents are saturated by an
infinite set of narrow meson states with the quantum numbers of
these currents, i.e. they can be represented in Euclidean space as follows
\ba
\label{cor1}
\Pi^J(Q^2)&=&\int d^4x\,e^{iQx}\langle\bar{q}\Gamma_J q(x)\bar{q}\Gamma_J q(0)
\rangle_{\text{planar}}\nl
&=&\sum_{n=0}^{\infty}\frac{Z_J(n)}{Q^2+m_J^2(n)}+\text{Subtractions}.
\ea
For instance, $\Gamma_V=\bm{\tau}\gamma_{\mu}$ for the vector isovector case,
where $\bm{\tau}$ are the Pauli matrices.
On the other hand, their high-energy asymptotics are given by
OPE~\cite{svz}. We will consider only the vector (V) and axial-vector (A)
isovector channels in the chiral limit.
After appropriate subtractions one has
\be
\label{opeva}
\Pi(Q^2)\xrightarrow[Q^2\rightarrow\infty]{}C_0\ln\!\frac{\mu^2}{Q^2}+
\sum_{k=1}^{\infty}\frac{C_k}{Q^{2k}}.
\ee
The coefficients $C_k$ only weakly depend on the scale $\mu$ and momentum $Q^2$
and can be calculated perturbatively.

Define
\begin{equation}
\label{trans}
\Pi_{\mu\nu}^{V,A}(Q^2)\equiv\left(-\delta_{\mu\nu}Q^2+Q_{\mu}Q_{\nu}\right)
\Pi^{V,A}(Q^2).
\end{equation}
The OPE reads as follows
\be
\Pi^{V,A}(Q^2)=\frac{N_c}{12\pi^2}
\ln\!\frac{\mu^2}{Q^2}
+\frac{\alpha_s}{12\pi}
\frac{\langle G^2\rangle}{Q^4}
+\frac{4\xi^{V,A}}{9}\pi\alpha_s
\frac{\langle\bar{q}q\rangle^2}{Q^6}+\mathcal{O}\left(\frac{1}{Q^8}\right),
\label{V}
\ee
where
$\xi^{V}=-7,\, \xi^{A}=11$.
The symbols $\langle G^2\rangle$ and $\langle\bar{q}q\rangle$ denote the
gluon and quark condensate respectively.
The residues are parametrized as follows
\begin{equation}
\label{resid}
Z_{V,A}(n)\equiv2F_{V,A}^2(n).
\end{equation}
with $F_{V,A}(n)$ being electromagnetic decay constants
(see~\cite{we2} for details).

The sum rules simply follow from comparison at each power of $Q^{-2}$ of
OPE~\eqref{V} with the sum in Eq.~\eqref{cor1} after
summing up over resonances (in a chiral invariant way)
and subtracting irrelevant infinite constants.

Consider the linear ansatz for the mass spectrum ($n=0,1,2,\dots$)
\be
\label{lin}
m^2_J(n)=an+m^2, \quad F^2_J(n)\equiv F^2=\text{Const}.
\ee
Such an ansatz is typically given by the strings, where
the universal slope $a$ is proportional to the string
tension $T$: $a=2\pi T$.
Up to an infinite constant one has
\be
\label{V2}
\Pi(Q^2)=
\sum_{n=0}^{\infty}\frac{2F^2}{Q^2+an+m^2}
=
-\frac{2F^2}{a}\,\psi\!\left(
\frac{Q^2+m^2}{a}\right)+\text{Const},
\ee
with $\psi$ being the Digamma function.
Let us introduce the notations
\be
z\equiv\frac{Q^2}{a},\qquad x\equiv\frac{m^2}{a}.
\ee

The $\psi$-function has an asymptotic representation at large argument
\be
\label{asymp}
\psi(z+x)=\ln{(z+x)}-\frac{1}{2(z+x)}-\sum_{k=1}^{\infty}\frac{B_{2k}}{2k(z+x)^{2k}},
\ee
Here $B_{2k}$ denote the Bernoulli numbers (see Appendix~A).
Expanding Eq.~\eqref{asymp} at $z\gg x$ one arrives at (see Appendix~B)
\be
\label{asymp2}
\psi(z+x)=\ln{z}-\sum_{k=1}^{\infty}\frac{(-1)^kB_k(x)}{kz^k},
\ee
where $B_k(x)$ are the Bernoulli polynomials. We could not find expansion~\eqref{asymp2}
in the mathematical handbooks, so a derivation is given in Appendix~B. This
expansion is quite remarkable. It completely determines OPE for the linear spectrum
through the Bernoulli polynomials and {\it vice versa}. Namely, equating expansions~\eqref{opeva}
and~\eqref{asymp2} one arrives at
\begin{equation}
\label{gensr}
\frac{2F^2}{a}\ln\!\frac{a}{Q^2}+2F^2\sum_{k=1}^{\infty}\frac{(-1)^ka^{k-1}B_k(x)}{kQ^{2k}}
+\text{Const}=\frac{N_c}{12\pi^2}\ln\!\frac{\mu^2}{Q^2}+\sum_{k=1}^{\infty}\frac{C_k}{Q^{2k}}.
\end{equation}
One can always make logarithms equal by an appropriate choice of Const. Then
\be
\label{F}
F^2=\frac{N_ca}{24\pi^2},
\ee
and the general structure of sum rules for the linear spectrum is
\be
\label{genstr}
C_k=\frac{(-1)^k N_c a^k B_k(x)}{12\pi^2k}.
\ee
Thus, the weak dependence on $\mu$ and $Q^2$ of the coefficients $C_k$ is replaced by
a constant in the case of the linear spectrum.

The sum rules considered in~\cite{we2,sh,gp,bean,sim1,afon,sim2,sh2,we,ar} are, in fact, particular cases of concise expression~\eqref{genstr}
for $k=1,2,3$ up to adding some separated resonances~\footnote{Similar expressions were derived also in~\cite{gp2} in a different way.
}.
We expect that one can trust these sum rules for $k\lesssim7$. The reason
is that the non-zero ({\it i.e.} even) Bernoulli numbers converge only up to $B_6$, see
Eq.~\eqref{bernnumb}. Then the divergence sets in reflecting the asymptotic nature of the
expansion.

Let us consider the vector case as an example. The first three sum rules~\eqref{genstr}
are (see Eq.~\eqref{V})
\ba
\label{2v}
0&=&x-1/2,\\
\label{3v}
\frac{\alpha_s\langle G^2\rangle}{12\pi F^2a}&=&x^2-x+1/6,\\
\label{4v}
-\frac{2\xi^{V}\pi\alpha_s\langle\bar{q}q\rangle^2}{3F^2a^2}&=&
x\left(x-1/2\right)\left(x-1\right),
\ea
where $F^2$ is given by Eq.~\eqref{F}. In the first sum rule, Eq.~\eqref{2v}, one
has zero in the l.h.s. due to the absence of local gauge-invariant dim2 condensate in OPE.

Consider the quantity in the l.h.s. of Eq.~\eqref{4v}, with $F^2$
being substituted from Eq.~\eqref{F},
\be
\label{devia}
\epsilon\equiv-\frac{16\pi^3\xi^{V}\alpha_s\langle\bar{q}q\rangle^2}{N_ca^3},
\ee
(we remind that although $\alpha_s$ and the quark condensate depend on the renormalization scale,
this dependence is almost negligible for the combination $\alpha_s\langle\bar{q}q\rangle^2$).
Below we will be interested in the solution $x=1/2$, corresponding
to $\langle\bar{q}q\rangle=0$. The quantity $\epsilon$ shifts that
solution,
\be
\label{shift}
x=\frac12+\delta.
\ee
Let us estimate how large is this shift due to a non-zero value of
quark condensate. From the Gell-Mann-Oakes-Renner relation,
\be
m_{\pi}^2f_{\pi}^2=-(m_u+m_d)\langle\bar{q}q\rangle,
\ee
with the standard inputs (the scale of the chiral symmetry breaking, $1\div1.2$ GeV is
assumed) $m_u+m_d\approx12$ MeV, $f_{\pi}\approx92.4$ MeV and $m_{\pi}\approx135$ MeV
one obtains $\langle\bar{q}q\rangle\approx-(235\,\text{MeV})^3$.
The QCD coupling constant at the same scale is
$\alpha_s\approx0.5$. The last
quantity we need is the slope $a$. If we accept an averaged value
$a\approx1.14$ GeV$^2$ from~\cite{bugg} then we obtain from
Eq.~\eqref{devia}: $\epsilon\approx0.07$.
Substituting Eq.~\eqref{shift} into the r.h.s. of Eq.~\eqref{4v}
and neglecting the term $\mathcal{O}(\delta^3)$, one arrives at
\be
\delta\simeq-4\epsilon.
\ee
It can be seen now that in the real world the corrections are
of the order of $\delta\approx -0.3$.

Our subsequent analysis is valid, strictly speaking, for small
corrections, $|\delta|\ll0.5$, {\it i.e.} either in the limit of
small enough quark condensate or in the limit of large enough
slope. The latter means that the ground state is largely separated
from the first radial excitation.
Nevertheless, one can hope that the analysis is relevant
for the real world: If one considers the planar limit of QCD (as we
do), one regards the corrections
$\mathcal{O}(1/N_c)=\mathcal{O}(1/3)$ as small and this turns out
to be a good approximation to the real world. In our case the
magnitude of corrections is comparable with that of the planar limit.
With this justification in mind, we will take $x=1/2$ as a starting
point in our analysis and will show that this works rather well for deriving
various qualitative results.

\section{Maximally dual spectrum}

It is well known that any linear spectrum is dual to the perturbation theory at
one loop with free quarks
since it reproduces the parton model logarithm in Eq.~\eqref{gensr}. However, in such a
perturbation theory the power corrections are absent, while one cannot avoid them saturating
correlators by the linear spectrum. In this situation it is interesting to find an ansatz
which minimizes the power corrections. We will refer to such an ansatz as
Maximally Dual Spectrum (MDS).

First of all we note that from the absence of dim2 condensate, corresponding to $C_1=0$
in Eq.~\eqref{genstr}, follows the solution $x=1/2$. Property~\eqref{pr2} yields
then $C_{2k+1}=0$. Thus, the half of power corrections is automatically zero.
In particular, $C_3\sim\alpha_s\langle\bar{q}q\rangle^2=0$, {\it i.e.} one has no CSB.
It is a good sign since the perturbation theory does not know about CSB.
Relation~\eqref{pr1} shows, however, that the zero values of $C_{2k+1}$ condensates correspond
to extremums of $C_{2k}$ ones and {\it vice versa}. Consequently, it is impossible to cancel
all power corrections. In QCD
this could mean that even if the chiral symmetry had not been
spontaneously broken ($\langle\bar{q}q\rangle=0$), some
nonperturbative condensates would have existed. Presumably, these
condensates are the purely gluon condensates with even powers of gluon fields, symbolically
$\langle G^{2k}\rangle$.
We point out the negative sign of the gluon condensate provided by Eq.~\eqref{3v} for $x=1/2$,
which indicates (if the quark-hadron duality is operative here)
that the obtained spectrum cannot be the spectrum of real
resonances, for modelling the latter one has to introduce some corrections.
The value of gluon condensate is, however, very sensitive to such corrections.
This fact can be used for
constructing the corrections to the masses and residues, which are consistent with
the physical values of OPE condensates~\cite{we2}. In the present
work we are interested in the ansatz, minimizing the power
corrections in the OPE. We will not bother about the sign of
condensates which it gives (as said above, it can be always improved by introducing small
corrections). Of our concern will be a usefulness of this ansatz for
a qualitative theoretical analysis.

Thus, minimizing the half of power corrections one inevitably maximizes the other half. Let us
formulate now the problem of minimization rigorously. For the time being it is convenient
to measure the condensates in Eq.~\eqref{genstr} in the units of slope $a$, so
$C_k\sim \frac{B_ k(x)}{k}$.
Consider the first $N$ sum rules. We define the maximally dual spectrum as an ansatz
providing the absolute minimum for the sum
\be
S(x)\equiv\sum_{k=1}^{N}\left(\frac{B_k(x)}{k}\right)^2.
\ee
Consequently, one should solve the equation
\be
\frac{dS(x)}{dx}=2\sum_{k=1}^{N}\frac{1}{k}B_k(x)B_{k-1}(x),
\ee
and substitute the solutions into
\be
\frac{d^2S(x)}{dx^2}=2\sum_{k=1}^{N}\left(B_{k-1}^2(x)+
\frac{k-1}{k}B_{k}(x)B_{k-2}(x)\right),
\ee
in order to to check the sign. Here relation~\eqref{pr1} was exploited.
The solution of the minimization problem turns out to depend on $N$.
For $N<16$ the minimum is delivered by $x=1/2$. For $N\geqslant16$ and uneven,
the minimum is the same, but for the even $N$ one obtains
$x\xrightarrow[N\rightarrow\infty]{}\left\{\frac14,\frac34\right\}$, {\it i.e.} the
roots of $B_{2k}(x)$ in asymptotics (see Eq.~\eqref{pr5}).
However, as pointed out above, we cannot
trust the sum rules at $N>7$ since the asymptotic divergence sets in.
For this reason the last term gives the main contribution to the sums above
since some $N$ and, hence, completely  defines their properties.
Thereby we
arrive at our corollary: {\it The maximally dual vector linear spectrum is}
\be
\label{mds}
m^2(n)=a(n+1/2).
\ee
This is nothing but a LS-type of spectrum for the vector states.
In fact, the same conclusion follows for any choice of minimality criterion of the kind
\be
S^r(x)\equiv\sum_{k=1}^{N}\left|\frac{B_k(x)}{k}\right|^r,\qquad r>0.
\ee

The maximally dual spectrum~\eqref{mds} is chirally symmetric in
the sense that the order parameters of CSB, $\langle\bar qq\rangle$ and
$f_{\pi}$, are automatically zero due to Eqs.~\eqref{2v}
and~\eqref{4v}. This is not surprising since the perturbation
theory is explicitly chirally symmetric.

\section{Corrections to spectrum}

The exactly linear spectrum is, of course, only approximation. In reality there are
corrections as to masses, $m^2(n)=a(n+x)+\Delta(n)$, as to residues $F^2(n)$. In the
real world a smooth function $\Delta(n)$ hardly exists. Even in a much simpler case,
dim2 QCD in the large-$N_c$ limit, the 't~Hooft model~\cite{dim2}, the spectrum is obtained
as a result of solution of the eigenvalue problem for a highly nontrivial integral
equation. As a result one has the linear spectrum only asymptotically, for the actual
masses one observes seemingly random fluctuations around the asymptotic spectrum. In
the real world one should have something similar. One can try to interpolate these
deviations by a smooth function $\Delta(n)$. There are arguments~\cite{we2} that
for the self-consistency of sum rules these deviations should then decrease in $n$ at
least exponentially. An alternative way to demonstrate this point is presented
in Appendix~C. However, the self-consistency of sum rules admits also the existence
of finite number of states, whose masses and residues are not subjected to a smooth
in $n$ parametrization. These states should be inserted "by hand" with the help of
$\delta$-function. This subject we consider in the present section.

Each separated state contributes to the sum rules as follows
\be
\frac{F^2_l}{Q^2+m^2_l}=F_l^2\sum_{j=0}^{\infty}\frac{(-1)^j m_l^{2j}}{Q^{2(j+1)}},
\ee
where index $l$ enumerates these states. Each individual contribution is, generally speaking,
quite large in comparison with typical quantities appearing in the sum rules from the OPE side.
On the other
hand, the linear ansatz is a good approximation. Thus, in order not to spoil the sum rules
it is reasonable to require that the sum over these contributions is approximately equal to
the same
sum over these states when they are subjected to the linear parametrization, {\it i.e.} we
require the duality between the actual spectrum and the linear one in the sense of approximate
coincidence of
their expansions at large Euclidean momentum. This approximation
we will denote by the sign "$\simeq$". Let us separate in this way the first $k$ states.
One then has
\be
\label{dev1}
\sum_{l=0}^{k}F_l^2m_l^{2i}\simeq\sum_{l=0}^{k}F^2(l+x)^i,\quad i=0,1,\dots.
\ee
It is convenient to parametrize the deviations from the linear spectrum through the dilatations,
\ba
\label{dil1}
F_l^2&=&\lambda_l F^2,\\
m^2_l&=&\Lambda_l a(l+x).
\ea
When $\lambda_l=\Lambda_l=1$ one has no
deviations. The requirement of duality~\eqref{dev1} reads then as follows
\be
\label{dev2}
\sum_{l=0}^{k}\lambda_l(\Lambda_l)^i(l+x)^i\simeq\sum_{l=0}^{k}(l+x)^i,\quad i=0,1,\dots.
\ee
For the axial-vector mesons the first sum rule in Eq.~\eqref{dev2} has an additional
pole at $Q^2=0$ coming from PCAC due to propagation of the massless pion. This sum rule reads,
\be
\frac{f_{\pi}^2}{F^2}+\sum_{l=0}^{k}\lambda_l^A\simeq k,
\ee
where $f_{\pi}=87$ MeV (in the chiral limit~\cite{gl}) is the weak pion decay constant.

Let us separate the first vector ($\rho$-meson) and axial-vector ($a_1$-meson) mesons and
assume that only these states can be strongly affected by CSB as their masses are close to
$\Lambda_{\text{CSB}}$.
We have then for the sum rules with $i=0,1$,
\ba
\label{srr1}
1&\simeq&\lambda_ {\rho},\\
\label{srr2}
1&\simeq&\frac{f_{\pi}^2}{F^2}+\lambda_{a_1},\\
1&\simeq&\lambda_ {\rho,a_1}\Lambda_ {\rho,a_1}.
\label{srr3}
\ea
In addition, these relations are supplemented with Eq.~\eqref{F} and the
solution for intercept, $x=1/2$, the latter means the duality between the
real spectrum and MDS.

In the real world with the chiral symmetry breaking there are some relations
between the spectral parameters of the lowest states which independently hold
in many realistic models and are in a good agreement with the experiment.
In essence, they are consequence of "one resonance + continuum" ansatz.
These relations are
\ba
\label{inp1}
F_{\rho}^2&\approx&2f_{\pi}^2,\\
\label{inp2}
m_{a_1}^2&\approx&2m_{\rho}^2,\\
m_{\rho}^2&\approx&\frac{24\pi^2}{N_c}f_{\pi}^2.
\label{inp3}
\ea
The first relation, Eq.~\eqref{inp1}, is the KSFR relation~\cite{ksfr}.
The second one was originally obtained by S.~Weinberg within the spectral sum rules~\cite{wein}
using Eq.~\eqref{inp1}.
Afterward it became clear that this formula expresses the fact of maximal mixing of
longitudinal component of $a_1$-meson with the pion after CSB,
while between the vector and scalar states this does not occur~\cite{gh}. Relation~\eqref{inp3}
was originally obtained within the Borel sum rules~\cite{mrho}. At present it is understood
as a consequence of vector meson dominance and usually it holds in models respecting this
property (see, {\it e.g.},~\cite{ar3}).

We also should respect relations~\eqref{inp1}-\eqref{inp3}. The standard point of view on the
sum rules is that they cannot completely determine the mass spectrum. This is related to the
fact that OPE is only asymptotic expansion for the correlators. The sum rules can give
only some relations. To get predictions one should supplement them by some external inputs.
First of all we observe that taking one of relations~\eqref{inp1}-\eqref{inp3} as input,
the other two relations are automatically reproduced by sum rules~\eqref{srr1}-\eqref{srr3},
giving as a byproduct the relation for the slope,
\be
\label{slope}
a\simeq2m_{\rho}^2\simeq\frac{48\pi^2}{N_c}f_{\pi}^2,
\ee
and for the axial-vector decay constant $F_{a_1}\simeq f_{\pi}$. For the parameters of dilatations
this reads: $\lambda_{\rho}\simeq\Lambda_{\rho}\simeq1$, $\lambda_{a_1}\simeq1/2$ and
$\Lambda_{a_1}\simeq2$. Thus, the mass and residue of $\rho$ meson are not affected by CSB
in our approximation.
It is important to note that the obtained slope coincides with the one given by the LS model,
Eq.~\eqref{ls1},
{\it i.e.} we have not just a LS-type spectrum in Eq.~\eqref{mds}, but exactly this spectrum!

However, in the axial-vector case the sum rules with $i>1$ are now violated. In particular,
at $i=2$ the contribution $2f_{\pi}^2m_{a_1}^4$ appears, giving rise to the
condensate $\pi\alpha_s \langle\bar{q}q\rangle^2\approx(4f_{\pi})^6$.
In this way we reproduce the conclusion of Ref.~\cite{knecht}:
{\it For QCD in the large-$N_c$ limit spontaneous CSB with $f_{\pi}^2\neq0$ necessarily
implies the existence of non-zero local order parameters}, where it was deduced from analysis
of finite number of resonances within the sum rules. In our case this conclusion follows
for an infinite number of states (which is more consistent with the large-$N_c$ limit),
with important relations~\eqref{inp1}-\eqref{inp3} being satisfied.

In principle, one could add more separate resonances and obtain a better agreement with
the experiment or accommodate the actual values of condensates. However, the theoretical
selfconsistency cannot be independently checked in this case because we do not know an analog
of Eqs.~\eqref{inp1}-\eqref{inp3} for, say, "2 resonances + continuum" ansatz. So we prefer
to stop at this stage.

We would like to point out that as a matter of fact we do not need any
inputs~\eqref{inp1}-\eqref{inp3}.
Due to PCAC they all can be reproduced automatically if we separate the ground axial-vector
meson.
Namely, if one neglects the exponentially small terms
and possible contributions from higher orders of perturbation theory~\cite{ijmp,sc},
the residues are given by the formula (see Appendix~C)
\be
\label{povres2}
F^2(n)=c\frac{dm^2(n)}{dn},
\ee
where the factor $c$ is determined from the coefficient in front of the perturbative logarithm,
Eq.~\eqref{F},
$c=N_c/(24\pi^2)$. Let us deviate the lowest state from the linear spectrum,
\be
\label{devi1}
m^2(n)=a\left[n+x+\varepsilon\delta(n)\right],
\ee
where $\delta(n)$ is the $\delta$-function. Since the derivative of linear function is equal
to the corresponding finite difference, one has from Eq.~\eqref{povres2} for small variations
\be
\label{devi2}
F^2(n)=c\left[m^2(n+1)-m^2(n)\right]\simeq ca\left[1-\varepsilon\delta(n)\right].
\ee
Thus, even without sum rules one can see that if the mass of the ground state is increased in
comparison with the value predicted by the linear spectrum, the decay constant of this particle
is decreased,
\be
\frac{\delta m^2(0)}{\delta F^2(0)}\simeq-\frac1c.
\ee
Now let us write the same for the axial-vectors in terms of dilatations.
The spectrum~\eqref{devi1} will be
\be
\label{asp}
m_{a_1}^2(n)=a\left(n+\frac12+\frac{\Lambda_{a_1}-1}{2}\delta(n)\right).
\ee
Keeping in mind Eq.~\eqref{devi2}, it leads to
\be
\label{conn}
\lambda_{a_1}\simeq1+\frac12(1-\Lambda_{a_1}).
\ee
The system of Eqs.~\eqref{conn} and~\eqref{srr3} has two solutions,
$\{\lambda_{a_1},\Lambda_{a_1}\}=\{1,1\},\{1/2,2\}$. The latter is valid only if there is
the additional contribution $1/2$ to the first
sum rule, Eq.~\eqref{srr2}. Identifying this contribution with the pion pole due to PCAC, all
the results above
follow immediately. Thus, the enhancement of the mass of axial-vector meson and the appearance
of pion pole
are intimately related. As a byproduct, Eq.~\eqref{conn} gives the constraint:
$m_{a_1}<\sqrt{3}\,m_{\rho}$.

\section{Sum rules with finite number of states}

In this section we digress from the main line and consider the case of sum rules with finite
number of resonances. There is a quite popular method of treating with such type of sum rules,
the so-called
Finite Energy Sum Rules (FESR) approach (for a review see, {\it e.g.},~\cite{raf}). In short,
the idea is to separate some (usually one) narrow resonances and make a cut-off at some energy,
replacing the rest by the perturbative QCD continuum. These sum rules usually work fairly well
for differences of correlators, say $\Pi^V(Q^2)-\Pi^A(Q^2)$, but for individual correlators they
are not so good. A reason is that replacing the contribution of infinite number of states
(any selfconsistent analysis of infinitely narrow mesons inevitably implies the large-$N_c$ limit,
hence, the infinite number of states) by a cut-off is a quite rough interpolation. As a result,
one has an uncontrollable growth of condensates, $C_k\sim\sum F^2(n)m^{2(k-1)}(n)$, which is
unrealistic to compensate by the choice of cut-off at $k>2$ in all sum rules simultaneously.
In practice, the condensates, say
$C_3\sim\alpha_s\langle\bar{q}q\rangle^2$, are much smaller than even the first term,
$\sim F^2(0)m^{4}(0)$ in the given case. Consequently, the realistic condensates can appear
only after some tricky cancellations of large numbers. Such predictions hardly can be stable
in response to variations of parameters.

We would like to propose another type of interpolation which is free from this drawback. Assuming
linearity of
spectrum (this is well settled experimentally, see, {\it e.g.},~\cite{ani}) one can sum over
resonances and arrive at sum rules~\eqref{genstr}. Then one can regard the r.h.s. of
Eq.~\eqref{genstr} as interpolating functions for contribution of finite number of states,
fitting the parameters $F^2$, $a$, and $x$. In particular, choosing a value for $x$ near the
roots of $B_3(x)$ naturally provides a small condensate $C_3$. Such sum rules are automatically
renormalized during the summation procedure, as a result the dependence on the cut-off disappears.
We expect that these sum rules should work at $k>2$ in Eq.~\eqref{genstr}, when the complete set
of roots of $B_k(x)$ on the interval $[0,1]$ sets in.

Let us demonstrate an example of using such sum rules. Consider the third sum rule,
\be
\label{tsr}
C_3(x)\sim x(x-1/2)(x-1),
\ee
and assume the one-resonance saturation. Since in practice $C_3(x)$ is small in comparison with
the typical values given by the r.h.s. of Eq.~\eqref{tsr} ({\it i.e.} given by the hadronic part)
this term can be introduced by small variations of $x$ near the roots of the r.h.s. of
Eq.~\eqref{tsr}: $x\rightarrow x+\varepsilon$. Then
\be
C_3(1/2+\varepsilon)\sim -\frac{\varepsilon}{4},\qquad
C_3(\{1,0\}+\varepsilon)\sim \frac{\varepsilon}{2}.
\ee
Thus
\be
\label{otn1}
\frac{C_3(1/2+\varepsilon)}{C_3(\{1,0\}+\varepsilon)}\sim-\frac12,
\ee
while from OPE, Eq.~\eqref{V}, one has
\be
\label{otn2}
\frac{C_3^V}{C_3^A}=\frac{\xi^V}{\xi^A}=-\frac{7}{11}.
\ee
We do not bother about the exact absolute value in Eq.~\eqref{otn1} since we have not varied $F^2$
and $a$.
What is important for us is the sign in Eqs.~\eqref{otn1} and~\eqref{otn2}. The negative sign is
possible only when $x_{\rho}\simeq1/2$ and $x_{a_1}\simeq1$ (we do not consider the unphysical
possibility $x_{a_1}\simeq0$).
Therefore, we obtain the Weinberg relation, Eq.~\eqref{inp2}, making use of the third sum rule
only. Thus, the sum rules under consideration work in the region where the usual ones fail for the
case of one separated resonance.
Combining this relation with the Weinberg sum rules or FESR at $k=1,2$ one arrives at the
statement which is converse to that of cited in the previous section:
{\it For QCD in the large-$N_c$ limit spontaneous CSB with non-zero local order parameters
necessarily
implies the existence of $f_{\pi}^2\neq0$} (\`a la Coleman-Witten theorem~\cite{cw}).

One can consider more resonances in Eq.~\eqref{tsr}, {\it i.e.} regard the r.h.s. as an
approximant for several states. This will result in the spectrum $m_{\rho}^2(n)\simeq a(n+1/2)$
and $m_{a_1}^2(n)\simeq a(n+1)$, which is a particular spectrum of the generalized LS amplitude,
the so-called
Ademollo-Veneziano-Weinberg (AVW) dual amplitude~\cite{avw}. We remind that the LS dual amplitude
is an amplitude for the reaction $\pi+\pi\rightarrow\pi+\pi$ with linearly rising Regge
trajectories and the correct chiral properties. It predicts
the following spectrum for the $\rho$-meson trajectory and its daughters in the chiral limit
\be
\label{ls1}
m^2_{\text{\tiny LS}}(n,J)=2m_{\rho}^2(n+J-1/2), \qquad J>0.
\ee
For $J=0$ one has to substitute $J=1$ in Eq.~\eqref{ls1}, {\it i.e.} the scalar and vector mesons
are degenerate.
The AVW amplitude is a generalization of the LS one to the reactions $\pi+A\rightarrow B+C$. It
contains also the spectrum for the pion trajectory (which includes the axial-vectors) and its
daughters, predicting
\be
\label{ls2}
m^2_{\text{\tiny AVW}}(n,J)=2m_{\rho}^2(n+J).
\ee
The experiment does not support the AVW spectrum~\cite{we3}:
It fails in describing the parity doubling on the daughter trajectories. Theoretically the reason
is clear:
The AVW spectrum leads to wrong chiral properties of correlators (the absence of chiral symmetry
restoration
at high energies, within the matching to effective models see~\cite{we5}). Thus, this spectrum
can be considered only as a low-energy approximation. For instance, it predicts a reasonable value
for the chiral constant $L_{10}$~\cite{gl},
\ba
\label{L10}
L_{10}&=&-\frac18\left[\Pi^V(0)-\Pi^A(0)\right]=\frac{\psi(x_V)-\psi(x_A)}{32\pi^2}\nl
&\approx&\frac{-1.96+0.58}{32\pi^2}\approx-4.4\cdot10^{-3},
\ea
which is close to the prediction of one-resonance + continuum ansatz,
\be
\label{L102}
L_{10}=-\frac14\left[\frac{F_{\rho}^2}{m_{\rho}^2}-\frac{F_{a_1}^2}{m_{a_1}^2}\right]
\simeq
\frac{-2+0.5}{32\pi^2}\approx-4.7\cdot10^{-3},
\ee
This happens because $L_{10}$ is saturated at low energies. But if a physical quantity includes
the high-energy asymptotics of correlators, for example, the electromagnetic pion mass difference
(for references see, {\it e.g.},~\cite{we4}),
\be
\label{dmp}
\Delta m_{\pi}=-\frac{3\alpha}{16\pi m_{\pi}
f_{\pi}^2}\int_0^{\infty}dQ^2Q^2\left[\Pi^V(Q^2)-
\Pi^A(Q^2)\right],
\ee
the AVW spectrum completely fails resulting in a divergence.

Thus, sum rules~\eqref{genstr} can be used for finite number of states, interpolating the hadronic contributions for $k>2$. They supplement FESR at $k=1,2$. Combining these two types of sum rules one can successfully describe the low-energy spectrum.

\section{Phenomenology and cluster duality}

The experimental spectrum of light nonstrange mesons is depicted on Fig.~1. The phenomenological
analysis which led to that plot is performed in~\cite{ej} (see also~\cite{lat}).
In short, the states below 1.9~GeV are
taken from the Particle Data~\cite{pdg}. Above 1.9~GeV only a few states in~\cite{pdg} are cited.
The
resonances are taken from the Crystal Barrel Collaboration data~\cite{bugg}. The main reason is
that
this is the only experiment which performed a systematic detailed study of the energy region
1.9-2.4~GeV. For reliability, only the states seen at least in two reactions were selected.

A prominent feature of the experimental spectrum is the existence of nearly equidistant
clusters of states with a growing number of resonances. Qualitatively this coincides with the
picture typically given by the strings, where the meson spectrum behaves as $m^2(n,J)\sim n+J$
($n$ is the number
of radial excitation and $J$ is the spin). Fixing the sum $n+J=N$ one obtains the $N$-th cluster
containing $N+1$ states. The chiral and axial symmetries of classical QCD Lagrangian, which are
expected to be approximately restored at high energies (parity doubling), multiply the number of
states in each cluster.

It is convenient to introduce the cluster spectrum, $M^2(N)=A(N+X)$, which parametrizes
the behavior of the experimental spectrum as a whole. The quantity $A$ has the physical
sense of mean slope and $AX$ does of mean intercept. These parameters can be easily calculated
from the experimental spectrum, they are presented in Table~1. We considered several cases.
First of all, the vector and axial-vector isovector mesons, the sum rules for these channels
are considered in the present work. Then all vector mesons, where the analysis is the same.
Finally, all light nonstrange mesons. Moreover, as the first cluster contains only two states,
it may be not well justified to consider this cluster on equal footing with the others, so we
also displayed in Table~1 the data with the first cluster excluded. Since the data of Crystal
Barrel
(the last two clusters) are still absent in Particle Data, we show also the results without this
data
(the first three clusters). It turns out that the results vary only a little.

\begin{center}
\begin{table}
\begin{tabular}{lccc}
\hline
$I^G,J^{PC}$ & Clusters & $A$, GeV$^2$ & $X$\\
\hline
$1^+,1^{--}$ and $1^-,1^{++}$ & 0-4 & 1.15 & 0.54 \\
$1^+,1^{--}$ and $1^-,1^{++}$ & 1-4 & 1.14 & 0.55 \\
$1^+,1^{--}$ and $1^-,1^{++}$ & 0-2 & 1.11 & 0.59 \\
$J=1$ & 0-4 & 1.12 & 0.53 \\
$J=1$ & 1-4 & 1.13 & 0.51 \\
$J=1$ & 0-2 & 1.10 & 0.56 \\
All mesons & 0-4 & 1.13 & 0.54 \\
All mesons & 1-4 & 1.13 & 0.55 \\
All mesons & 0-2 & 1.14 & 0.54 \\
\hline
\end{tabular}
\caption{Parameters of cluster spectrum $m^2(N)=A(N+X)$, $N=0,1,2,3,4$
in Fig.~\eqref{fig}.}\label{tab:table1}
\end{table}
\end{center}

\vspace{-1.3cm}

A striking feature of experimental spectrum is that on average it behaves as
MDS, with the slope being very close to the predicted one, Eq.~\eqref{slope}.
Since MDS is chirally symmetric, we conclude that
{\it although CSB shifts significantly the masses of some states, when one considers the whole
light nonstrange meson spectrum
with more than 80 resonances, i.e. when the statistical weight of these several states is small,
one observes the approximately chirally symmetric spectrum, as if CSB were absent at all!}
The experimental spectrum also reveals the universality of linear spectrum for any quantum numbers,
which seems to allow the generalization of the conclusions for vector channels made in this paper
to arbitrary ones. It must be emphasized that Particle Data (the first three clusters) contains
enough states to arrive at the conclusion above. The Crystal Barrel Data convincingly confirms
this observation.

How the existence of approximately equidistant clusters is related
to the hypothesis of duality of real spectrum to MDS?
If deviations from MDS are sporadic to both sides, then the duality naturally leads to the clusters
with the positions determined by MDS. However, we should take into account the possibility of
systematic
deviations from MDS to one particular side, which should be compensated by deviations to opposite
side
in other channels in order to provide a stable cluster. This can be achieved by generalization of
duality considered above in the following way
\be
\label{gend}
\Pi^J(Q^2)=\frac1k\sum_{l=1}^k\Pi^J_{(l)}(Q^2),
\ee
where $l$ enumerates all possible sets of quantum numbers with fixed spin $J$. The spin defines a
Lorentz tensor structure in front of the correlator, like in Eq.~\eqref{trans} for $J=1$.
This tensor structure is extracted in Eq.~\eqref{gend}.
The mean correlator
in the l.h.s. of Eq.~\eqref{gend} is supposed to be saturated by MDS, hence, defining the
positions
of clustering the poles in the r.h.s. This generalized duality we would call the {\it cluster
duality}.
The case considered in Section~IV is a particular case of cluster duality for two
channels,
when the spectrum of one of them (vector mesons) is locally dual to MDS ({\it i.e.} coincides).
Then Eq.~\eqref{gend}
requires at least the global duality to MDS for the axial-vector states.

\section{Dimension-two condensate}

The authors of Ref.~\cite{z1} advocated the idea that although the local dim2 gauge invariant
operator is absent in the standard OPE in the chiral limit, one can construct a gauge
non-invariant dim2 operator from the gluon fields and the minimal value of its vacuum
expectation value (achieved in the Landau gauge), $\lambda^2\sim\langle G\rangle$, could
have a definite physical sense, encoding some important nonperturbative effects. This
subject soon got a certain popularity (see, {\it e.g.}, references in~\cite{ar}). Usually one
attempts
to improve the agreement with phenomenology introducing this condensate and in such a way
one obtains an estimate on its value. In particular, within the sum rules in the large-$N_c$
limit some estimates have been derived in~\cite{ar,sim2} (in~\cite{we2} it was noticed, however,
that
within this theoretical setting the dim2 condensate hardly can be detected). Let us
pursue this way, {\it i.e.} let us assume the existence of dim2 gluon condensate and try to
estimate its value within the linear spectrum.

Consider the vector channel. The dim2 gluon condensate modifies the first sum rule~\cite{z2},
Eq.~\eqref{2v}
\be
\label{2vg}
-\frac{\alpha_s\lambda^2}{4\pi^3}\simeq2F^2(x-1/2).
\ee
The axial-vector channel will have the same contribution, but the appearance of $f_{\pi}^2$
hampers the estimation in that case. As we have shown (see Table~1), in reality
$\Delta x\equiv x-1/2>0$.
Since phenomenologically $\lambda^2<0$ ("effective tachyonic gluon mass") we have agreement
in the sign. Suppose that the deviation from the maximally dual ansatz, $x=1/2$, happens
completely due to existence of $\lambda^2\neq0$ (this yields disagreement in the sign for the
quark
condensate term in the third sum rule, Eq.~\eqref{4v}, but for the time being we do not bother
about that). We expect that this should give a significant overestimation for the quantity
in question. Using Eq.~\eqref{F} one arrives at
\be
\label{2vges}
-\frac{\alpha_s\lambda^2}{\pi}\simeq\frac{N_c}{3}\,a\Delta x.
\ee
Substituting the approximate experimental values for $x$ and $a$ from Table~1 into
Eq.~\eqref{2vges} one obtains the estimate $-\frac{\alpha_s\lambda^2}{\pi}\approx0.03\div0.06$
GeV$^2$
(we neglect the extremal values).
This supposedly overestimated value turns out to be less up to one order of magnitude
than ones, usually cited in the literature. In fact, the experimental deviations $\Delta x$
have the same smallness as the chiral corrections. Say, in the LS amplitude the pion
mass yields the correction
\be
x_{\text{\tiny LS}}=\frac12+\frac{m_{\pi}^2}{m_{\rho}^2-m_{\pi}^2}\approx0.53.
\ee
Thus, in the chiral limit we cannot make the case in estimating quantitatively the dim2
condensate within our theoretical setting, let alone the fact that the chiral corrections are
usually less than $\mathcal{O}(1/N_c)$ ones. As a matter of fact,
our estimate is well consistent with the absence of dim2 condensate.

\section{Conclusions}

We give a new insight into the phenomenological success of the Lovelace-Shapiro dual amplitude:
Its spectrum is maximally dual to the perturbation theory among the linear spectra. This spectrum
turns out to describe the positions of clusters of meson states near some values of energy.
This indicates on the existence of certain duality between the real spectrum and the maximally
dual one. The reason seems to be the chiral symmetry restoration
in the excited states since the maximally dual spectrum
corresponds to zero order parameters of the chiral symmetry
breaking in QCD.
In the given paper we have demonstrated how this duality may work within the sum rules
giving quite encouraging results. Our analysis shows that there is a hope to describe
quantitatively the global properties of observed light meson spectrum within the QCD sum rules,
at least with the accuracy typical for the large-$N_c$ limit.

We would like to mention also that the QCD sum rules suggest a possible way to prove the
asymptotic linearity of the light meson spectrum, at least for the vector channels. Namely,
due to relation~\eqref{povres} this is tantamount to an independent demonstration that the
decay constants asymptotically tend to a constant value.

A possible continuation of the present work is an analysis of channels with arbitrary spins.
The experimental spectrum  shows that the results must be similar.



\section*{Acknowledgments}
I would like to thank M. A. Shifman, A. I. Vainstein and V. I. Zakharov for very useful
discussions during the Workshop "QCD and String Theory", July 2-14, Benasque, Spain.
The work was supported by
CYT FPA, grant 2004-04582-C02-01, CIRIT GC, grant 2001SGR-00065,
RFBR, grant 05-02-17477, grant LSS-5538.2006.2, and by Ministry of Education and Science
of Spain.


\section*{Appendix A. Bernoulli numbers and polynomials}

The Bernoulli numbers can be represented as follows
\ba
\label{ber}
B_{2k}&=&(-1)^{k-1}\frac{2(2k)!}{(2\pi)^{2k}}\zeta(2k),\\
B_{2k+1}&=&0, \quad k>0.
\label{unevenB}
\ea
Here $\zeta$ is the Riemann function. The first few numbers are
\begin{align}
\label{bernnumb}
B_0&=1,\quad B_1=-\frac12,\quad B_2=\frac16,\nl
B_4&=-\frac{1}{30},\quad B_6=\frac{1}{42}, \quad B_8=-\frac{1}{30}.
\end{align}

The Bernoulli polynomials are defined as
\be
\label{berpol}
B_k(x)=\sum_{m=0}^{k}\binom{k}{m}B_mx^{k-m}.
\ee
The first few polynomials are
\begin{align}
B_0(x)&=1,\quad B_1(x)=x-\frac12,\quad B_2(x)=x^2-x+\frac16,\nl
B_3(x)&=x^3-\frac32x^2+\frac12x=x(x-1/2)(x-1),\nl
B_4(x)&=x^4-2x^3+x^2-\frac{1}{30}.
\label{ffp}
\end{align}
The Bernoulli polynomials have many interesting properties. Below we cite those
which are relevant for our analysis. For $k>0$ one has
\begin{align}
\label{pr1}
B_k'(x)&=kB_{k-1}(x),\\
\label{pr2}
B_{2k+1}(0,1/2,1)&=0,\\
\label{pr3}
B_{2k}(0,1)&=B_{2k},\\
\label{pr4}
B_{2k}(1/2)&\xrightarrow[k\rightarrow\infty]{}-B_{2k},\\
\label{pr5}
B_{2k}(x)&\xrightarrow[k\rightarrow\infty]{}\sim\cos(2\pi x),\\
\label{pr6}
B_{2k+1}(x)&\xrightarrow[k\rightarrow\infty]{}\sim\sin(2\pi x).
\end{align}
It should be noted that in fact asymptotic properties~\eqref{pr4}-\eqref{pr6} set in
quite rapidly. Say, $B_2(\frac12)=-\frac12B_2$, but already $B_4(\frac12)\approx-0.97B_4$.
As to Eq.~\eqref{pr5}, we are interested in the roots on the interval~$[0,1]$ only.
Already for $B_2(x)$ the approximate roots 0.21 and 0.79 are close to the asymptotic ones,
$\frac14$ and $\frac34$.

\section*{Appendix B. Expansion of {\large\bm{$\psi$}} function in Bernoulli polynomials}

In this Appendix we derive Eq.~\eqref{asymp2}. Using property~\eqref{unevenB}
one can cast Eq.~\eqref{asymp} into a form
\be
\label{A1}
-\psi(z+x)=-\ln{z}-\ln{\left(1+\frac{x}{z}\right)}+
\frac{1}{2(z+x)}
+\sum_{k=2}^{\infty}\frac{B_{k}}{k(z+x)^{k}},
\ee
Expand the last three terms in the r.h.s. of Eq.~\eqref{A1} at $z\gg x$,
\be
\label{A2}
-\ln{\left(1+\frac{x}{z}\right)}=\sum_{k=1}^{\infty}\frac{(-1)^k}{k}\frac{x^k}{z^k}
=\sum_{k=1}^{\infty}\frac{(-1)^k}{kz^k}\binom{k}{0}B_0x^{k-0},
\ee
\be
\label{A3}
\frac{1}{2(z+x)}=\frac{1}{2z}\sum_{k=0}^{\infty}(-1)^k\frac{x^k}{z^k}
=\sum_{k=1}^{\infty}\frac{(-1)^k}{kz^k}\binom{k}{1}B_1x^{k-1},
\ee
\ba
\label{A4}
\sum_{k=2}^{\infty}\frac{B_{k}}{k(z+x)^{k}}&=&
\sum_{k=2}^{\infty}\frac{B_k}{kz^k}\sum_{l=0}^{\infty}(-1)^l\binom{k+l-1}{l}\frac{x^l}{z^l}\nl
&=&\sum_{k=2}^{\infty}\frac{1}{z^k}\sum_{m=2}^{k}\frac{(-1)^{k-m}}{m}\binom{k-1}{k-m}B_m x^{k-m}\nl
&=&\sum_{k=2}^{\infty}\frac{(-1)^k}{kz^k}\sum_{m=2}^{k}\binom{k}{m}B_m x^{k-m}.
\ea
In the last case we multiplied the sums according to the formula
\be
\sum_{n=2}^{\infty}a_ny^n\sum_{k=0}^{\infty}b_k(n)y^k=
\sum_{n=2}^{\infty}y^n\sum_{k=2}^{n}a_kb_{n-k}(n),\nonumber
\ee
then used identity $\binom{k-1}{k-m}=\frac{m}{k}\binom{k}{m}$ and relation~\eqref{unevenB}.
Summing Eqs.~\eqref{A2}, \eqref{A3}, \eqref{A4} and substituting the result into Eq.~\eqref{A1} one
obtains finally
\be
-\psi(z+x)=-\ln{z}+\sum_{k=1}^{\infty}\frac{(-1)^k}{kz^k}\sum_{m=0}^{k}\binom{k}{m}B_m x^{k-m}.
\ee
Using definition of Bernoulli polynomials~\eqref{berpol} one arrives at Eq.~\eqref{asymp2}.

\section*{Appendix C. Systematic nonlinear corrections to spectrum}

Let us analyze the constraints on the possible nonlinear corrections to the string-like spectrum.
A naive way is to expand in powers of corrections in the sum of Eq.~\eqref{V2}. However,
the validity of this procedure requires a strong convergence properties and {\it apriori}
we do not know them. Instead we will make use of the Euler-Maklaurin summation formula, which is
an asymptotic series for a sum
\ba
\label{em}
\sum_{n=0}^{N}f(n)&=&\int_{0}^{\infty}f(x)dx-B_1\left[f(0)+f(N)\right]\nl
&+& \sum_{k=2}^{\infty}\frac{B_k}{k!}\left[f^{(k-1)}(N)-f^{(k-1)}(0)\right].
\ea
We are interested in functions with the property
$f^{(k)}(N)\xrightarrow[N\rightarrow\infty]{}0$, $k=0,1,\dots$, for which
Eq.~\eqref{em} reads
\be
\label{em2}
\sum_{n=0}^{N}f(n)=\int_{0}^{\infty}f(x)dx-
\sum_{k=1}^{\infty}\frac{B_k}{k!}f^{(k-1)}(0).
\ee
In our case
\be
f(n)\sim\frac{F^2(n)}{z+n+x+\Delta(n)}.
\ee
For the linear spectrum, $F^2(n)=\text{const}$, $\Delta(n)=0$, expression~\eqref{em2}
gives the asymptotic expansion of Digamma function, Eq.~\eqref{asymp}. In particular, the
logarithm
appears from the integral. This imposes a constraint on $F^2(n)$: To reproduce the
analytical structure of OPE (i.e. to have the series in $z^{-k}$ only) the residues
has to behave as
\be
\label{povres}
F^2(n)\sim\frac{dm^2(n)}{dn},
\ee
up to some exponentially small terms~\cite{we2}.
We will not be interested in the latter. The problem is reduced to the corrections to the linear
mass spectrum,
\be
\label{residred}
f(n)\sim\frac{1+\Delta'(n)}{z+n+x+\Delta(n)}.
\ee
Substituting Eq.~\eqref{residred} into Eq.~\eqref{em2} and expanding in inverse powers
of $z$ one obtains an infinite series at each power because now $\Delta^{(k)}(0)\neq0$
at $k>1$. Say, at $1/z$ one has a contribution
\be
\label{diverg}
\frac{1}{z}:\quad\sim\sum_{k=1}^{\infty}\frac{B_k}{k!}\Delta^{(k-1)}(0).
\ee
Now we require that small corrections do not have to generate uncontrollable divergences in
all results. If we want to preserve the present scheme of analytical treatment with the
sum rules we need to impose the convergence of the sums like~\eqref{diverg}.
From relation~\eqref{ber} for large even $k$ one has then the following condition
in the case of sum~\eqref{diverg}
\be
\label{conv}
\frac{1}{2\pi}\left|\frac{\Delta^{(k)}(0)}{\Delta^{(k-1)}(0)}\right|<1.
\ee
For power-like corrections, $\Delta(n)\sim n^{-\alpha}$ ($\alpha>0$), condition~\eqref{conv}
cannot be fulfilled, as well as for the corrections like $\Delta(n)\sim n^{\alpha}e^{-\beta n}$
($\beta>0$), where $\alpha$ is not a positive integer. But for purely exponential ones,
$\delta(n)\sim e^{-\beta n}$,
condition~\eqref{conv} can be satisfied if $\beta<2\pi$. The same is true for a special
type of "mixed" corrections: $\Delta(n)\sim n^{\alpha}e^{-\beta n}$ ($\alpha,\beta>0$),
where $\alpha$ is integer. At other powers of $1/z$ the
situation is more involved. Let us give only estimate for the exponential corrections
(they have a property $\Delta^{(k)}(0)=\left(\Delta'(0)\right)^k$),
\be
\label{diverg2}
\frac{1}{z^m}:\quad\sim\sum_{k=1}^{\infty}\frac{B_k(c_m)^k}{k!}\Delta^{(k-1)}(0),
\ee
where $c_m=\mathcal{O}(1)$, $c_m>1$. One obtains then $\beta<2\pi/c_m$ for convergence.

Thus, within the present scheme there are strong constraints on the derivatives
of corrections. The power-like corrections cannot satisfy them, but the exponential
ones do can. On the other hand, the exponent $\beta$ is also constrained. Actually,
it can be so small that in the beginning of spectrum the decreasing is slower
than, say, $1/n$. The presented estimates are in agreement with the results and fits
obtained in~\cite{we2}.

\begin{center}
\begin{figure*}
\vspace{-7cm}
\hspace{-3cm}
\includegraphics[scale=0.9]{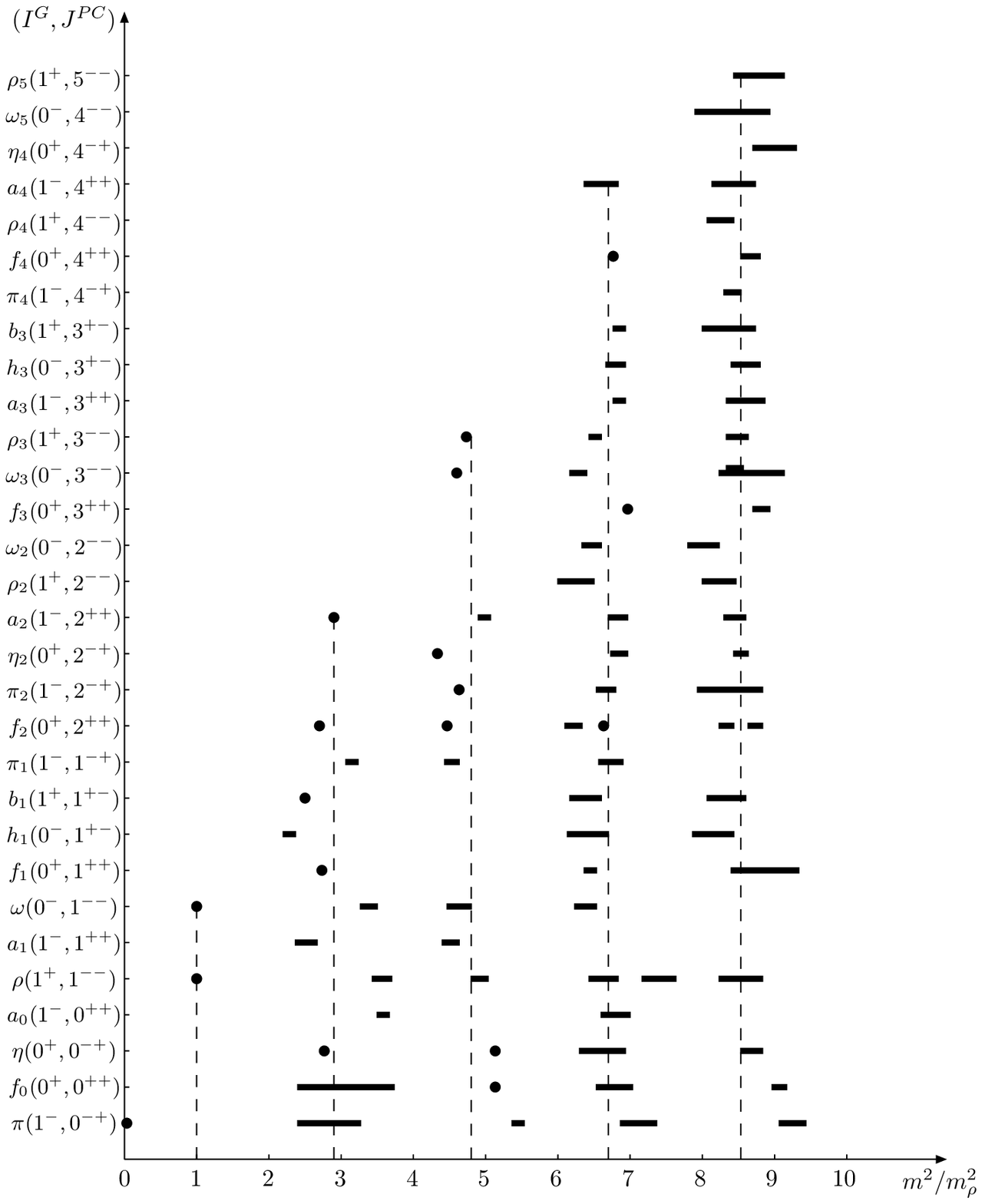}
\vspace{-7.5cm}
\caption{\label{fig}\small The spectrum of light non-strange mesons
from~\cite{pdg} and~\cite{bugg} (for the last two clusters) in
units of $m_{\rho(770)}^2$. Experimental errors are indicated.
Circles stay when errors are negligible. The dashed lines mark the
mean (mass)$^2$ in each cluster. The clustering occurs at
0.78, 1.32, 1.70, 2.00, and 2.27 GeV. The following states are
displayed (in MeV):\quad $\pi$: $140$, $1300\pm100$, $1812\pm14$,
$2070\pm35$, $2360\pm25$;\quad $f_0$: 
$1200-1500$,
$1770\pm12$, $2020\pm38$, $2337\pm14$;\quad $\eta$: 
$1294\pm4$, $1760\pm11$, $2010^{+35}_{-60}$, $2285\pm20$;\quad $
a_0$: 
$1474\pm19$, $2025\pm30$;\quad $\rho$:
$775.8\pm0.5$, $1465\pm25$, $1720\pm20$, $2000\pm30$, $2110\pm35$,
$2265\pm40$;\quad
$a_1$: $1230\pm40$, $1647\pm22$
;\quad $\omega$: $782.59\pm0.11$, $1400-1450$, $1670\pm30$,
$1960\pm25$;\quad $f_1$: $1281.8\pm0.6$, $1971\pm15$,
$2310\pm60$;\quad $h_1$: $1170\pm20$, $1965\pm45$,
$2215\pm40$;\quad $b_1$: $1229.5\pm3.2$, $1960\pm35$,
$2240\pm35$;\quad $\pi_1$: $1376\pm17$, $1653^{+18}_{-15}$,
$2013\pm25$;\quad $f_2$: $1275\pm1$, $1638\pm6$, $1934\pm20$,
$2001\pm10$, $2240\pm15$, $2293\pm13$;\quad $\pi_2$: $1672\pm3$,
$2005\pm15$, $2245\pm60$;\quad $\eta_2$: $1617\pm5$, $2030\pm16$,
$2267\pm14$;\quad
$a_2$: $1318.3\pm0.6$, $1732\pm16$, $2030\pm20$, 
$2255\pm20$;\quad $\rho_2$: $1940\pm40$, $2225\pm35$;\quad
$\omega_2$: $1975\pm20$, $2195\pm30$;\quad $f_3$: $2048\pm8$,
$2303\pm15$;\quad $\omega_3$: $1667\pm4$, $1945\pm20$,
$2255\pm15$, $2285\pm60$;\quad $\rho_3$: $1688\pm2.1$, $1982\pm14$,
$2260\pm20$;\quad $a_3$: $2031\pm12$, $2275\pm35$;\quad $h_3$:
$2025\pm20$, $2275\pm25$;\quad $b_3$: $2032\pm12$, $2245\pm50$;
\quad $\pi_4$: $2250\pm15$;\quad $f_4$:
$2018\pm6$, $2283\pm17$;\quad $\rho_4$: $2230\pm25$;\quad $a_4$:
$2005^{+25}_{-45}$, $2255\pm40$;\quad $\eta_4$: $2328\pm38$;\quad
$\omega_5$: $2250\pm70$;\quad $\rho_5$: $2300\pm45$.}
\end{figure*}
\end{center}

\end{document}